\documentclass[12pt,a4paper]{article}

\usepackage{epsfig,graphicx}
\usepackage{amsmath,amsfonts,amssymb}
\usepackage{citesort}

\newcommand{\unit}{\leavevmode\hbox{\small1\kern-3.6pt\normalsize1}}

\def\lsim{\raise0.3ex\hbox{$\;<$\kern-0.75em\raise-1.1ex\hbox{$\sim\;$}}}
\def\gsim{\raise0.3ex\hbox{$\;>$\kern-0.75em\raise-1.1ex\hbox{$\sim\;$}}}

\begin{document}

\thispagestyle{empty}
\begin{flushright}
FTUAM-12-84\\
IFT-UAM/CSIC-12-19\\
{March 2012}
\end{flushright}

\begin{center}
{\Large \textbf{ 
Indirect Dark Matter Searches and Models
} 
}

\vspace{0.5cm}
\hspace*{-1mm}
Carlos Mu\~noz\\[0.2cm] 
\textit{Departamento de F\'{\i}sica Te\'{o}rica, Universidad Aut\'{o}noma de Madrid\\
Cantoblanco, E-28049 Madrid, Spain}\\[0pt]
\textit{Instituto de F\'{\i}sica Te\'{o}rica UAM/CSIC, Universidad Aut\'{o}noma de Madrid\\
Cantoblanco, E-28049 Madrid, Spain}\\[0pt]

\begin{abstract}
Indirect dark matter searches are briefly reviewed.
Current experimental data from satellites and Cherenkov telescopes searching for antimatter and gamma rays in galactic and extragalactic regions, are compared with predictions from theoretical models of dark matter. The analysis is focused on WIMPs such as the neutralino and the sneutrino, and a superWIMP such as the gravitino, in several interesting supersymmetric models.
In particular, the discussion is carried out in the context of $R$-parity conserving models such as the MSSM, the NMSSM, and an extended NMSSM, and the $R$-parity violating model $\mu\nu$SSM.
\end{abstract}
\end{center}

\vspace*{3cm}
\hspace{5cm} Neutralino, Sneutrino, Gravitino

\newpage

\section{Introduction}
\label{Introduction}

Elucidating the nature of the dark matter (DM) left over from the Big Bang and its possible detection constitutes a key challenge in modern physics \cite{reviewmio}. 
Evidence indicating the presence of DM can be obtained at 
very different scales: from cosmological ones through the analysis of the 
angular anisotropies in the cosmic microwave background (CMB) radiation, down to 
galactic scales considering lensing and galaxy dynamics studies. 
Although the amount of DM has been determined with huge precision by 
WMAP through the measurements of the CMB \cite{Komatsu:2010fb}, 
\begin{equation}
\Omega_{DM} h^2 = 0.1109\pm 0.0056\ , 
\label{relic}
\end{equation}
its composition is still unknown beyond the fact that it has to be mainly non-baryonic. 
Since 
within the standard model of particle physics there are 
no viable non-baryonic candidates,  
the existence of DM represents one of the most compelling evidences for 
physics beyond the standard model \cite{reviewmio}.

A new particle with the following properties is needed: 

\noindent {\it i)} It must be stable or long-lived, because it must have been produced after the Big Bang and still present today.

\noindent {\it ii)} It must be neutral, because otherwise it would bind to nuclei and would be excluded from unsuccessful searches for exotic heavy isotopes.

\noindent {\it iii)} It must reproduced the observed amount of DM (\ref{relic}).

Actually, a particle with weak interactions and a mass of the order of the electroweak scale, the so-called 
WIMP (Weakly Interacting Massive Particle) \cite{reviewmio}, is able to fulfill property {\it iii)}. 
The reason is that the relic density of WIMPs can be computed with the result
\begin{equation}
\Omega_{\mbox{\tiny WIMP}}  
h^2
\simeq \frac{3\times 10^{-27}\ \mathrm{cm^3\ s^{-1}}}
{\langle\sigma_{ann}\ v \rangle}\ , 
\label{relicdensity}
\end{equation}
where the denominator is the annihilation cross section of two DM particles averaged over their
velocity
distribution. Then, if a new particle with weak interactions exists in Nature, its annihilation cross section 
turns out to be of the right order,
${\langle\sigma_{ann}\ v \rangle}\sim 3\times 10^{-26}$ cm$^3$ s$^{-1}$,
to reproduce the observed density of the Universe (\ref{relic}).

The forthcoming onset of the large hadron collider (LHC) will provide information about the nature of particle physics at the electroweak scale. The LHC could produce a new kind of particle with a mass of the order of the GeV-TeV that could be in principle a candidate for DM. Such a production and detection would be of course a great success, but not a complete test of the DM theory. 
Even if we are able to measure the mass and interactions of the new particle, checking 
whether the observations are fulfilled (\ref{relic}), we would never be able to test if the
candidate is stable on cosmological scales.
A complete confirmation can only arise from experiments where the DM particle is detected as part of the galactic halo or extragalactic structures. As we will discuss below, this can come from
direct and indirect DM searches.
Actually, there has been an impressive progress on this issue in recent years, with significant improvements in the precision and 
sensitivity of experiments.
The combination of LHC data with those provided by direct and indirect searches can be a crucial tool for the identification of the DM.
Thus, these three detection strategies are ideal because they allow exploring in a complete way many different particle DM models. Not only that, in the case of a redundant detection (in two or more different experiments) the combination of their data can provide good insight into the nature of the DM, maybe even allowing its identification.


The DM could be detected directly in underground laboratories through its elastic interaction with nuclei inside detectors. 
Actually, there are claims about DM signals by direct detection experiments such as DAMA/LIBRA \cite{dama}, CoGeNT \cite{cogent}, and CRESST \cite{cresst}. These claims seem to be consistent with a low-mass particle, possibly of about 10 GeV. However, other experiments like 
CDMS \cite{cdms}, XENON \cite{xenon} and SIMPLE \cite{simple} do not confirm this result.
Thus the situation is controversial.

In this work we will concentrate on indirect DM searches.
These are carried out in neutrino and Cherenkov telescopes, and satellites, through the analysis of the DM annihilation or decay products in the Sun, galactic center, galactic halo or extragalactic structures. Such products can be neutrinos, gamma rays and antimatter. 
In the next sections we will review the current experimental situation concerning
indirect searches of WIMPs through gamma rays and antimatter. 
Actually, claims by several authors about detection,
using {\it Fermi} satellite data from the galactic center and galaxy clusters,
have also appeared recently in the literature \cite{hooper0,frenk}.
As for the case of direct detection signals, the situation is also controversial.
As we will discuss below, there is no confirmation of detection by the
{\it Fermi} collaboration
or by other authors analyzing the data.
We will also review the comparison between the experimental data and the predictions from theoretical models containing WIMPs.
Although the zoo of DM candidates is huge \cite{reviewmio}, we will concentrate on supersymmetric candidates such as the neutralino \cite{reviewmio} and the sneutrino \cite{sneutrino}. Finally, we will discuss that the gravitino is an interesting (superWIMP) candidate for DM in $R$-parity breaking models, and
can in principle be detected through its decay products, such as gamma rays.

The paper is organized as follows.
In Section \ref{section 2} we will discuss four interesting supersymmetric models 
where the neutralino, sneutrino and gravitino are candidates for DM.
In Section \ref{section 3} we will review indirect WIMP searches through antimatter and gamma rays, and the implications for the theoretical models containing neutralinos and sneutrinos.
We will discuss how the DM is searched in the galactic halo, galactic center, dwarf spheroidal galaxies and clusters of galaxies.
We will also comment recent claims about possible DM detection in the galactic center and the Virgo cluster.
In Section \ref{section 4} a similar analysis will be carried out for superWIMP searches
through gamma rays, focusing on gravitino DM in $R$-parity breaking models.

\section{Models}
\label{section 2}

Supersymmetry (SUSY) is one of the most attractive theories for physics beyond the standard model, and actually, 
one of the main goals of the LHC is to find its signatures.
As mentioned in the Introduction, SUSY candidates for DM exist and we will discuss them
in the context of four interesting SUSY models.

\subsection{MSSM}

The Minimal Supersymmetric Standard Model (MSSM) is the most popular SUSY extension of the standard model. 
It has been studied in great detail in the
literature, 
and
it is the simplest extension: no extra fields are
included apart from the SUSY partners of the standard model fields. 
In addition to the Yukawa couplings for quarks and charged leptons, the
MSSM superpotential contains the so-called $\mu$-term
involving the Higgs doublet
superfields, $\hat H_1$ and $\hat H_2$,
\begin{equation}
\begin{array}{rcl}
W  = 
\epsilon_{ab}\ \mu  \hat H_1^a \hat H_2^b\ ,
\end{array}\label{superpotentialmssm}
\end{equation}
where 
$a,b$ are
$SU(2)$ indices.
The presence of the $\mu$-term  
is essential to avoid 
the appearance of an unacceptable Goldstone boson  
associated to a global $U(1)$
symmetry.
It is also necessary to generate chargino masses, and present experimental bounds
imply that $\mu$ must be larger than about 100 GeV.

On the other hand,
the MSSM superpotential 
conserves a discrete symmetry called 
$R$-parity (+1 for particles and -1 for superpartners), and therefore
SUSY particles are produced 
or destroyed only in pairs. As a consequence, the lightest 
supersymmetric particle (LSP) 
is absolutely stable, and therefore fulfills property {\it i)} described in the Introduction.
This implies that the LSP 
is a possible candidate for DM. 
It is remarkable that in interesting regions of the parameter space
of the MSSM
the LSP is the lightest neutralino, 
a physical superposition of the Bino, and neutral Wino and Higgsinos.
The neutralino is
obviously 
an electrically neutral particle, fulfilling therefore property {\it ii)}.
It is also a WIMP, fulfilling property {\it iii)}.
Thus, in the MSSM, the lightest neutralino 
is a very good
DM candidate \cite{reviewmio}.
Let us finally remark that the fact that the LSP is stable,
and typically neutral, implies that
a major signature in accelerator experiments 
for $R$-parity conserving models is represented by events with missing 
energy.

The phenomenological analysis of the MSSM can be carried out in two frameworks. One of them consists
of defining the soft SUSY-breaking terms at the grand unification scale,
$M_{GUT}$, and through the renormalization group equations (RGEs) to study the low-energy theory.
If the soft terms are assumed for simplicity universal at $M_{GUT}$, the model is usually called the
Constrained Minimal Supersymmetric Standard Model ({\bf CMSSM}).
Another possibility is to define the parameters of the MSSM directly at the electroweak scale, without any constraint from the RGEs. In this case the model is usually called the
effective MSSM ({\bf effMSSM}).



Unfortunately, the $\mu$-term  introduces a naturalness problem in the theory,
the so-called $\mu$ problem \cite{Kim:1983dt}. Note to this respect that
the $\mu$-term is purely SUSY, and therefore the natural
scale of $\mu$ would be $M_{GUT}$ or $M_{Planck}$.
Thus, any complete explanation of the electroweak scale must
justify the origin of $\mu$, i.e. why its value is of order $M_W$ and
not $M_{GUT}$
or $M_{Planck}$.
%
There are very interesting 
solutions to this problem that
necessarily introduce new structure beyond the MSSM at low energies. 
Several of these
solutions, and the associated SUSY models, are discussed below.

\subsection{NMSSM}

The Next-to-Minimal Supersymmetric Standard Model (NMSSM) 
provides an elegant solution to the $\mu$ problem of the MSSM via the
introduction of a singlet superfield $\hat S$ under the standard model gauge group. 
Substituting now the $\mu$-term in (\ref{superpotentialmssm}) by
\begin{equation}
\begin{array}{rcl}
W  = 
 \epsilon_{ab}\ \lambda \hat S \hat H_1^a \hat H_2^b\ + k \hat S\hat
S\hat S\ ,
\end{array}\label{superpotentialnmssm}
\end{equation}
%
when the scalar component of the superfield $\hat S$,
denoted by $S$, acquires
a vacuum expectation value (VEV) of order the SUSY breaking scale, 
an effective interaction $\mu \hat H_1 \hat H_2$ is generated
through the first term in  (\ref{superpotentialnmssm}), with  
$\mu\equiv\lambda \langle S\rangle$.
This
effective coupling is naturally of order the electroweak scale if the SUSY  breaking
scale is not too large compared with $M_W$. In fact, the NMSSM is the simplest
SUSY extension of the standard model in which the electroweak scale
exclusively originates from the SUSY breaking scale. 
The second term in (\ref{superpotentialnmssm}) 
is allowed by all symmetries,
and avoids, as the $\mu$-term in the MSSM, 
the presence of a Goldstone boson.

Due to the presence of the superfield $\hat S$,
in addition to the MSSM fields, the NMSSM contains an extra CP-even and CP-odd
neutral Higgs bosons, as well as one additional neutralino. These new fields
mix with the corresponding MSSM ones, giving rise to a richer and more complex
phenomenology.
For example, the results concerning the 
possible detection of neutralino DM turn out to be modified 
with respect to those of the MSSM in some regions of the parameter space.

\subsection{An extended NMSSM 
}
\label{extended}

An interesting extension of the NMSSM can help
us to explain
the origin of neutrino masses.
Since experiments induce us to introduce
right-handed neutrino superfields, $\hat \nu^c$, in the superpotential (\ref{superpotentialnmssm}), this 
can be extended with \cite{kitano}:
\begin{equation}
\begin{array}{rcl}
\delta W  = 
\ \epsilon_{ab} 
Y_\nu^{ij} \, \hat H_2^b\, \hat L^a_i \, \hat \nu^c_j 
+ \kappa^{ij} \hat S \hat \nu^c_i\hat \nu^c_j\ ,
\end{array}\label{deltasuperpotentialMajorana2}
\end{equation}
where $i,j$ are generation indices.
Here Majorana masses for right-handed neutrinos of the order of the electroweak scale are generated dynamically
through the VEV of the singlet $S$, $M_{\nu}=\kappa \langle S\rangle$. This is an example of 
a seesaw at the electroweak scale.
Light masses are then obtained with a value 
$m_{\nu} = Y_{\nu}^2 v_2^2/M_{\nu}$, 
which implies Yukawa couplings $Y_{\nu}$ of the order $10^{-6}$, i.e. of the same order than the electron Yukawa.


The left-handed sneutrino in the MSSM, even if it is the LSP, is not a viable DM candidate. Given its sizable 
coupling to the $Z$ boson, left-handed sneutrinos either annihilate too rapidly,
resulting in a very small relic abundance, or give rise to a large
scattering cross section and are excluded by direct DM
searches.

However, in this model a purely right-handed sneutrino can be a
  viable candidate for DM \cite{sneutrino}.
  Through the direct coupling to the singlet, the
  sneutrino cannot only be thermal relic DM reproducing
the WMAP result (\ref{relic}), but
  also have a large enough scattering cross section with nuclei to
  detect it.

\subsection{$\mu\nu$SSM}

As mentioned above, experiments induce us to introduce
gauge-singlet neutrino superfields.
Then, given the
fact
that sneutrinos are allowed to get VEVs,
we may wonder why not to use terms
of the type $\hat \nu^c \hat H_1\hat H_2$ 
to produce an effective  $\mu$ term.
This would allow us to solve the $\mu$ problem of the MSSM, 
without having to introduce an extra singlet superfield
as in case of the NMSSM.
This is the basic idea of the so-called `$\mu$ from $\nu$'
Supersymmetric Standard Model ($\mu$$\nu$SSM) \cite{LopezFogliani:2005yw,reviewsmunu}: natural particle content 
without $\mu$ problem.



In addition to the MSSM Yukawa couplings for quarks and charged leptons, 
the
$\mu$$\nu$SSM superpotential contains: 
%
\begin{equation}
W  = 
\ \epsilon_{ab}
Y_\nu^{ij} \, \hat H_2^b\, \hat L^a_i \, \hat \nu^c_j 
 + \epsilon_{ab} \lambda^{i} \, \hat \nu^c_i\,\hat H_1^a \hat H_2^b
+
\kappa^{ijk} 
\hat \nu^c_i\hat \nu^c_j\hat \nu^c_k\ .
\label{superpotentialmunussm}
\end{equation}
%
When the scalar components of the superfields $\hat\nu^c_i$,
denoted by $\tilde\nu^c_i$, acquire
VEVs of order the electroweak scale, 
an effective interaction $\mu \hat H_1 \hat H_2$ is generated
through the second term in  (\ref{superpotentialmunussm}), with  
$\mu\equiv
\lambda^i \langle \tilde \nu^c_i \rangle$.
The third type of terms in (\ref{superpotentialmunussm}) 
is allowed by all symmetries,
and avoids the presence of a Goldstone boson associated to a global $U(1)$
symmetry, similarly to the case of the NMSSM.
In addition, it contributes to generate 
effective Majorana masses for neutrinos at
the electroweak scale
$\kappa \langle \tilde \nu^c \rangle$.
Thus, the $\mu$$\nu$SSM solves the $\mu$ problem and explains the origin of neutrino masses 
by simply introducing right-handed neutrinos.

The above terms in the superpotential
produce the explicit breaking of R-parity (and lepton number)
in this model. The size of the breaking can be 
easily understood if we realize that in the limit where neutrino Yukawa couplings $Y_{\nu}$ are vanishing, the 
$\hat \nu^c$ are 
ordinary singlet superfields like the $\hat S$ of the 
NMSSM (\ref{deltasuperpotentialMajorana2}), 
without any connection with neutrinos,
and this model would be like the
NMSSM 
(with three singlets),
where R-parity is conserved.
Once we switch on the $Y_{\nu}$,
the 
$\hat \nu^c$ become right-handed neutrinos, and, as a consequence, R-parity
is broken. 
This breaking has to be small because of the
electroweak scale seesaw implying small values for $Y_{\nu}\sim 10^{-6}$.

Since
$R$-parity 
is broken, this means that the 
phenomenology of the $\mu$$\nu$SSM is going to be very different from the one 
of the MSSM/NMSSM.
Needless to mention, the LSP is no longer stable, and therefore the neutralino or the sneutrino, having very short lifetimes, are no longer viable candidates for DM.


On the other hand, let us suppose that the gravitino is the LSP. Since it has an interaction term in the Supergravity Lagrangian with the photon and the photino, and the latter 
and the left-handed neutrinos are mixed due to the breaking of $R$-parity, the gravitino will be able to 
decay into a photon and a neutrino. 
The decay is supressed both by  
the gravitational interaction (Planck mass) and by
the small $R$-parity violating coupling, thus
the lifetime of the gravitino can
be much longer than the age of the Universe, fulfilling condition
{\it i)}.
Additionally, adjusting the reheating temperature one can
reproduce the correct relic density (\ref{relic}) for each possible value of the gravitino 
mass (see e.g. \cite{Choi:2009ng} and references therein). Thus condition {\it iii)} can also be fulfilled.
As a conclusion, the gravitino, which can be classified as a superWIMP given its extremely weak interactions, is an interesting decaying DM candidate in $R$-parity violating models
\cite{yamaguchi}.

Since the gravitino decays producing a monochromatic photon with an energy half of the gravitino mass,
the prospects for detecting these $\gamma$ rays in satellite experiments can be very interesting, as we will discuss in Section \ref{section 4}.

















\section{Indirect WIMP Searches}
\label{section 3}

There are promising methods for the indirect detection of WIMPs by looking for evidence of their annihilations through anomalous
cosmic rays (CRs) produced in the galactic center, galactic 
halo or extragalactic structures such as dwarf spheroidal galaxies and clusters \cite{reviewindirect}.
These annihilations will produce $\gamma$ rays or antimatter, and fluxes of these particles can be measured in {\bf space-based detectors} such as {\it Fermi}-LAT ($\gamma$ rays) and PAMELA or AMS (antimatter). $\gamma$ rays can also be measured in {\bf Cherenkov telescopes} such as MAGIC, HESS or VERITAS.
Besides, 
{\bf neutrino telescopes} might detect 
WIMPs passing through the Sun and/or Earth.
They may be slowed below escape velocity by elastic scattering.
Then, the annihilation of WIMPs accumulated due to gravitational effects produces energetic neutrinos
that can be 
detected in underground (Super-Kamiokande), underwater (ANTARES) and under-ice (IceCube)
experiments.
In the following we will concentrate on satellite detectors and Cherenkov telescopes.




\subsection{Antimatter}

The observation of an excess of antiparticles 
with respect to the astrophysical background, could then be a signature of DM annihilations. 
Actually, PAMELA has measured the positron fraction, $e^+/(e^+ + e^-)$, up to 100 GeV, obtaining
an excess of positrons that increases with energy above 10 GeV \cite{pamela}.
Recently, {\it Fermi} \cite{fermipositrons} has found that the positron fraction increases with energy between 20 and 200 GeV, consistent with results reported by PAMELA.
However, there are several problems with a DM explanation for this large flux. First of all, an excess of antiprotons should have also been observed, but this was not the case.
Besides, PAMELA data would imply
$\sigma_{ann}\ v \sim 10^{-23}$ cm$^3$ s$^{-1}$, but, following (\ref{relicdensity}), this means that the value of $\Omega_{DM} h^2$ is much smaller than 0.1.
To avoid this problem one would have to require boost factors provided by clumpiness in the DM distribution ranging between
$10^{2}$ and $10^{4}$.
However, the high-energy positrons mainly come from a region within
{\it few kpc from the Sun} (those far away lose their energies during the propagation), where
boost factors larger than 10 are not expected.
On the other hand, astrophysical explanations for this excess are possible. For example, contributions to the fluxes of positrons from pulsars or CRs interacting with
giant molecular clouds \cite{pame,reviewindirect}.


\subsection{$\gamma$ rays}

An excess of $\gamma$ rays with respect to the astrophysical background could also be a signature of DM particle annihilations. As will be discussed below, searches for this excess can be carried out in different regions of the Milky Way or in extragalactic objects.

\subsubsection{Intermediate galactic latitudes}

\begin{figure}[t]
\begin{center}
\includegraphics[width=0.5\textwidth]{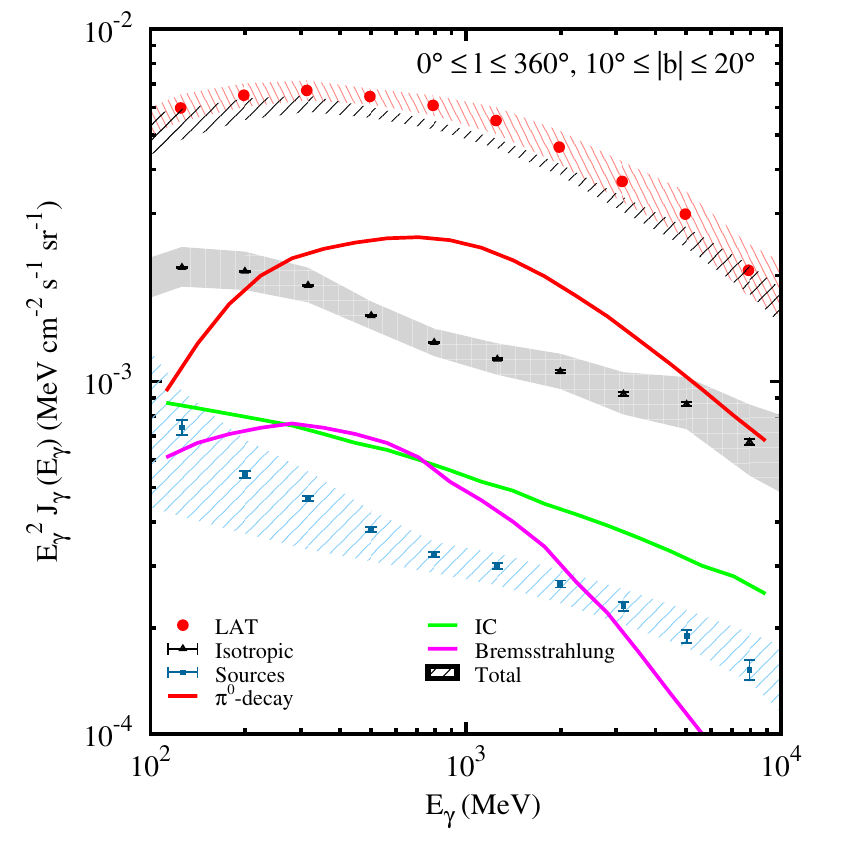}
\caption{\label{midlatitude} 
LAT data of diffuse emission intensity averaged
over all galactic longitudes for
latitude range $10^o\leq |b| \leq 20^o$ compared
with model, source, and UIB components.
Model (lines): $\pi^0$-decay, red; bremsstrahlung, magenta; IC, green.
Shaded/hatched regions: UIB, grey/solid; source, blue/hatched; total (model + UIB + source),
black/hatched. Figure from \cite{fermimid}.}
\end{center}
\end{figure}

The diffuse galactic $\gamma$-ray emission (DGE) is produced by CRs, mainly protons and electrons, interacting with the interstellar gas, via $\pi^0$-production and bremsstrahlung, and radiation field, via inverse Compton (IC) scattering. 
Measurements by the EGRET satellite, which covers the energy range 30 MeV to 30 GeV, indicated an excess of $\gamma$-ray emission $\geq 1$ GeV in all directions on the sky \cite{egret}.
DM explanations were proposed to solve this discrepancy. 
However, 5-month measurements for energies 100 MeV to 10 GeV and intermediate galactic latitudes,
$10^o\leq |b| \leq 20^o$, reported by {\it Fermi}-LAT,
which is conducting an all-sky $\gamma$-ray survey in the 20 MeV to $>$ 300 GeV
energy range,
show no excess \cite{fermimid}.
Fig.~\ref{midlatitude} from \cite{fermimid}
compares the LAT spectrum with the spectra of an {\it a priori} DGE model, and a point source contribution and
unidentified background (UIB) component derived from fitting the LAT data.
Overall, the agreement between the LAT-measured spectrum and the model shows that fundamental
processes are consistent with the data.

\subsubsection{Galactic center}

Another interesting possibility is to search for the DM particles in the galactic center, since it is expected to contain the largest density of DM within the Milky Way.
A preliminary analysis of the {\it Fermi}-LAT observations of the 
galactic center was reported in \cite{aldo}. 
The bulk of the $\gamma$-ray emission from that region was
explained with the detected sources and the DGE model.
Nevertheless, an unmodeled excess was present about 2--5 GeV.
The conclusion was that any attempt to disentangle a potential DM signal will require
deep understanding of the conventional astrophysics background.
An excess might be due to other astrophysical sources (for instance unresolved point sources).
Analyses within the {\it Fermi} collaboration are still under-way, and no 
further publications have appeared.

On the other hand,
utilizing three years of data from {\it Fermi}-LAT, it has been claimed that the spectrum of the $\gamma$-ray emission from the galactic center 
shows evidence of a spatially extended component which peaks at energies between 300 MeV and 10 GeV \cite{hooper0}, and that if interpreted as DM annihilations products, the DM particles with the correct relic density (\ref{relic}) should have a mass in the range of 7--12 GeV (if annihilating dominantly to leptons) or 25--45 GeV (if annihilating dominantly to hadronic final states).
Note that the former of the above mass ranges is consistent with signals reported by the direct detection experiments DAMA/LIBRA, 
CoGeNT,
and CRESST.
Let us finally remark that this result seems to be consistent with a cuspy density profile $\rho (r)\sim r^{-1.25}$ to $r^{-1.40}$. Although this kind of profiles are not obtained in dark-matter-only simulations, where a NFW density profile is typically obtained with a behavior $\rho (r)\sim r^{-1}$, it is worth noticing that when baryons are included in the analysis of the inner galactic region, the compression effect turns out to be important and can produce $\rho(r)\sim r^{-1.45}$, implying 
large $\gamma$-ray
fluxes 
\cite{adiabatic}.

However,
as mentioned above when discussing {\it Fermi}-LAT observations of the galactic center, a potential problem of the result obtained in \cite{hooper0} is that the conventional astrophysics background in the galactic center is not well understood.
In particular, it has been claimed that this emission might be consistent with a millisecond pulsar population in the central stellar cluster \cite{abazajian} or with 
CR effects \cite{cosmic}.
Also, in ref.~\cite{boya}, using a different spectrum of the point source at the galactic center
assumed by \cite{hooper0},
no DM particles are needed to explain the data.

To constrain particle physics models, one can compare the observations with 
the theoretical computation of the flux. The
observed differential flux at the Earth  
coming from a direction forming an angle
$\psi$ with respect to the galactic center is
\begin{equation}
\Phi_{\gamma}(E_{\gamma}, \psi)
=\frac{1}{2}\frac{\langle\sigma_{ann}\ v\rangle}{4 \pi m_{\mbox{\tiny DM}}^2}
\sum_i 
\frac{dN_{\gamma}^i}{dE_{\gamma}} B_i
  \int_{l.o.s.} \rho^2
\ dl\ ,
\label{flux}
\end{equation}
\noindent 
where the discrete sum is over all DM annihilation
channels,
$dN_{\gamma}^i/dE_{\gamma}$ is the differential $\gamma$-ray yield, $B_i$ is the branching ratio,
$\langle\sigma_{ann}\ v\rangle$ is the annihilation cross section averaged over its
velocity
distribution, $m_{DM}$ is the mass of the DM particle,
and $\rho$ is the assumed DM density in the galaxy. 
The integral is computed along the line of sign (l.o.s.) in the direction $\psi$.
Thus the result is factorized into the 'astrophysical factor' given by the integral which depends on the DM density, and the 'particle physics factor' in front which depends on the DM particle properties.

Recently, {\bf neutralino DM in the CMSSM} 
was studied \cite{ellis} using the above eq.~(\ref{flux}) and the 
current {\it Fermi}-LAT data. The conclusion was that the latter are not sensitive to any of the CMSSM scenarios with appropriate relic density 
studied.
This analysis was carried out assuming NFW and Einasto profiles.
In \cite{david},
a very light {\bf right-handed sneutrino DM in the extended NMSSM} discussed in 
Subsection \ref{extended}, was analyzed. In particular the possible detection of a sneutrino of about 10 GeV, compatible with results by DAMA/LIBRA, CoGeNT and CRESST, through its
$\gamma$-ray annihilation product from the galactic center, was discussed in detail. Assuming NFW, Einasto, and isothermal profiles, the conclusion is that
the fluxes are too small to be observed. Therefore, without a significant improvement
of the understanding of the background, one cannot constrain the relevant parameter space
of these models.

\subsubsection{Dwarf spheroidal galaxies}

Local group dwarf spheroidal galaxies (dSphs) are attractive targets because they are nearby, are largely DM dominated systems, and they are relatively free from $\gamma$-ray emission from other astrophysical sources since have no active star formation or  detected gas content. 
Thus, although their expected number of signal counts is smaller than the one from the galactic center,
given that they are further away, dSphs exhibit a favorable signal to noise ratio.  
However, 11-month measurements of 14 dSphs reported by 
{\it Fermi}-LAT show no significant $\gamma$-ray emission above 100 MeV \cite{dwarf}.

To constrain particle physics models, a sample of 8 dSphs without large uncertainties in the DM content was used. For each galaxy 
the collaboration modeled the DM distribution via a NFW density profile (neglecting boosts due to substructures in order to be conservative). With this information and using eq.~(\ref{flux}), upper limits on photon 
fluxes and on $\langle\sigma_{ann}\ v \rangle$ were derived as a function of the WIMP mass, for each dSph and for specific
annihilation channels \cite{dwarf} (see also \cite{conrad}).
Concerning the latter, continuum $\gamma$ rays produced by the decay of neutral
pions generated in the cascading of annihilation products is the origin 
of the fluxes. Typical products are quark-antiquark pairs.
An interesting case motivated by SUSY DM is a
$b\bar b$ final state.
Another final state motivated by SUSY, with a smaller branching fraction, is
$\tau^+ \tau^-$. 
An intermediate case with a mixed 
$b\bar b$ and $\tau^+ \tau^-$ in the final state is also
very common in SUSY.
The resulting integral flux above 100 MeV turns out to be at a level
below about $10^{-9}$ photons cm$^2$ s$^{-1}$
Of course, these results apply not only to neutralinos but to any model with this kind of final states.

\begin{figure}[t]
\begin{center}
\includegraphics[width=0.5\textwidth]{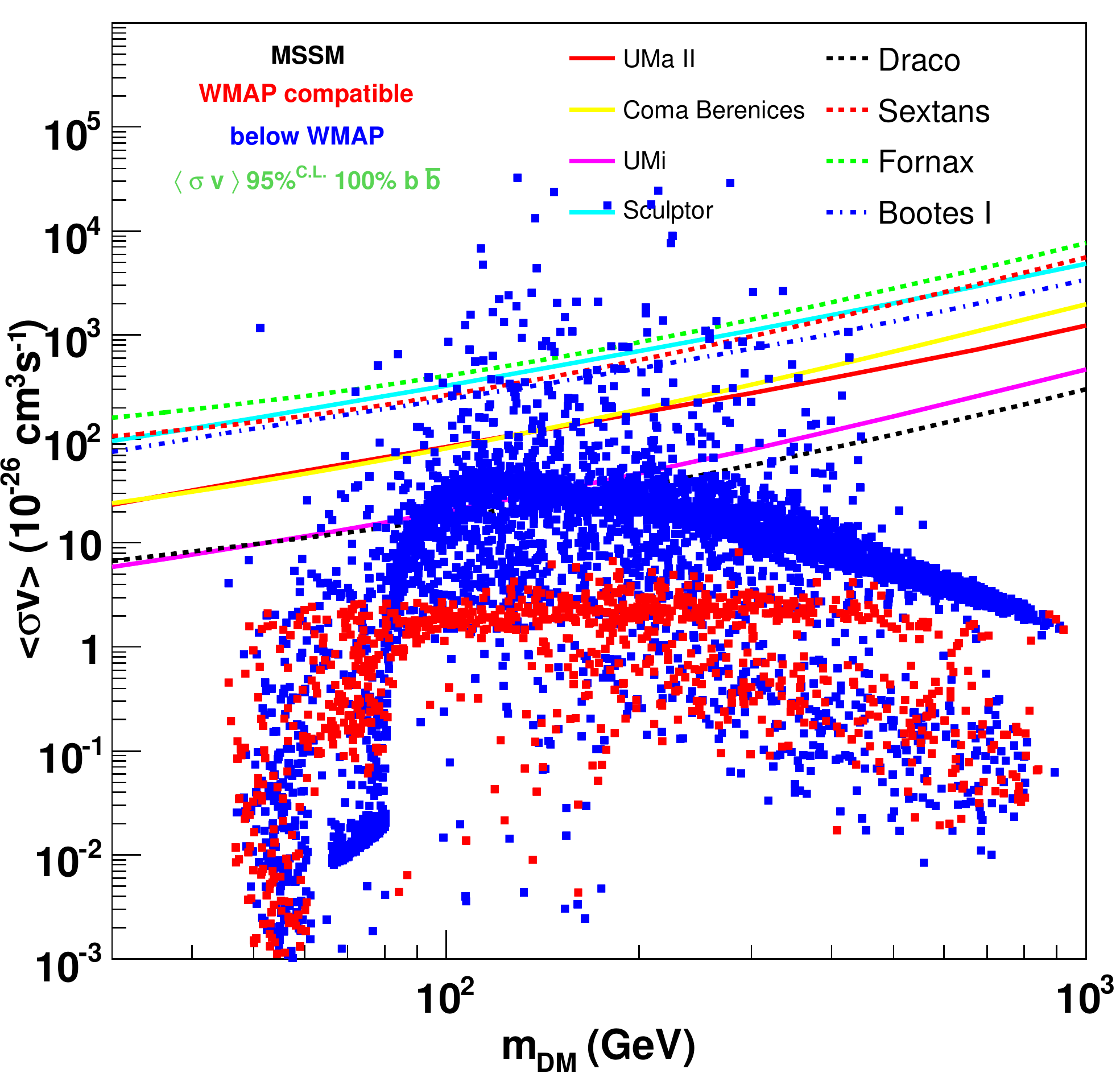}
\caption{\label{fig:one} effMSSM 
in the ($\langle\sigma_{ann}\ v \rangle$, $m_{\mbox{\tiny DM}}$) plane.
All points are consistent with all accelerator constraints and red points have
a neutralino thermal relic abundance consistent with WMAP.
Blue points have a lower thermal relic density but it is assumed that neutralinos still 
comprise all of the DM in virtue of additional non-thermal production processes.
The line indicate the {\it Fermi} 95$\%$ upper limits obtained from likelihood analysis
on the 8 selected dwarfs. Figure from \cite{dwarf}.}
\end{center}
\end{figure}

In Fig.~\ref{fig:one} from ref.~\cite{dwarf},
the LAT sensitivity 
in the ($\langle\sigma_{ann}\ v \rangle$, $m_{\mbox{\tiny DM}}$) plane
is compared with predictions from {\bf neutralino DM in the effMSSM}.
Draco and Ursa Minor dSphs set the best limits.
As expected, for neutralinos fulfilling (\ref{relic}), 
$\langle\sigma_{ann}\ v \rangle$ is given by 
$\sim 3\times 10^{-26}$ cm$^3$ s$^{-1}$ 
or by smaller values when coannihilation effects in some regions of the
parameter space are included.
We can see in Fig.~\ref{fig:one} that these (red) points remain unconstrained.

No excess has been observed either from dSphs in Cherenkov telescopes like HESS, VERITAS, MAGIC and Whipple, implying limits from these studies that vary between a few times
$\sim 10^{-23}$ to a few times $10^{-22}$ cm$^3$ s$^{-1}$ for a 1 TeV mass neutralino. Let us remark that Cherenkov telescopes are more sensitive to DM particles with high masses (higher than about 200 GeV), and their searches are thus complementary to those of {\it Fermi}.

\begin{figure}[t]
\begin{center}
\includegraphics[width=0.5\textwidth]{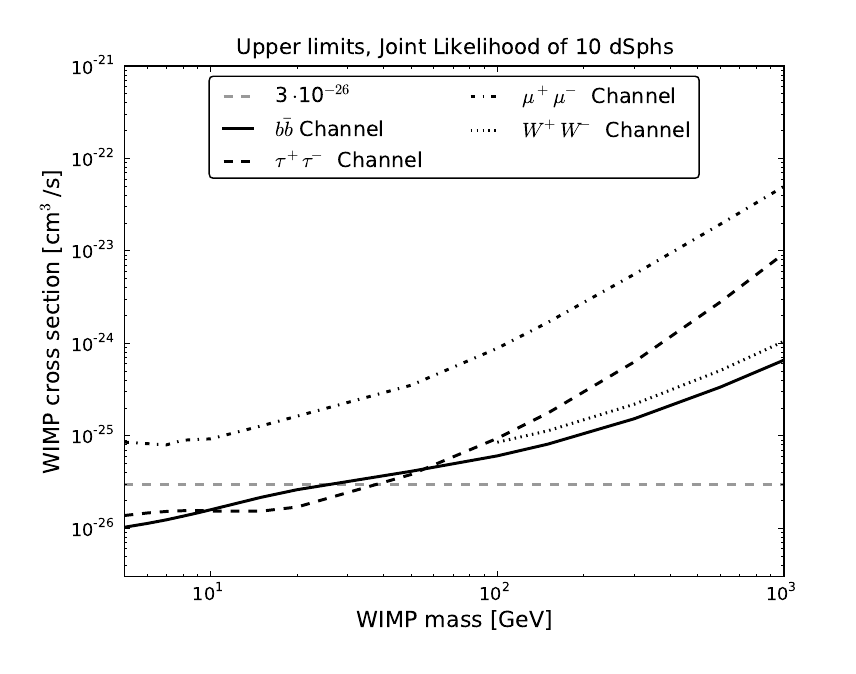}
\caption{\label{stack} Derived 95\% C.L. combined upper limits of the 10 selected dwarfs on a WIMP 
annihilation cross section for the $b\bar b$
channel, the $\tau^+\tau^-$ channel, the $\mu^+\mu^-$ channel, and the  $W^+ W^-$ channel.
The thermal cross section 
$\sim 3\times 10^{-26}$ cm$^3$ s$^{-1}$ is plotted as reference.
Figure from \cite{fermidwarfnew}.}
\end{center}
\end{figure}

In a recent work \cite{fermidwarfnew}, using 24 months of data, adding
Segue 1 and Carina to the sample of 8 dSphs analyzed in \cite{dwarf}, and including
the uncertainty in the DM distribution,
{\it Fermi}-LAT collaboration was able to obtain stronger constrains combining all
the dSph observations into a single joint likelihood function.
The upper limits on the annihilation cross section 
can be seen in Fig.~\ref{stack} from ref.~\cite{fermidwarfnew}.
Thus WIMPs with thermal cross sections are ruled out up to a mass of about 27 GeV
for the $b\bar b$ channel and up to a mass of about 37 GeV for the $\tau^+\tau^-$ channel.

\subsubsection{Clusters of galaxies}

Nearby clusters of galaxies are also attractive targets. 
They are more distant (the flux decreases with the cluster distance $D$ like $1/D^2$), but more massive than dSphs. Like dSphs, they are
very DM dominated. Besides, they typically lie at high galactic latitudes where the contamination from galactic $\gamma$-ray background emission is low. However,
11-month measurements of the 6 galaxy clusters, AWM 7, Fornax, M49, NGC 4636, Centaurus, and Coma, reported by 
{\it Fermi}-LAT show no 
excess \cite{clusters}.
Assuming a NFW profile for the DM density distribution of the clusters, 
the {\it Fermi} collaboration have explored the implications of the non-detection of DM
in terms of constraints on particle models, such as e.g. {\bf neutralino DM in the
effMSSM}.
In general, they turn out to be weaker than those found for dSphs.
Although they improve significantly the constraints obtained from
observations of Coma and Perseu clusters by HESS and MAGIC, respectively.

Recently, 8 galaxy clusters assuming a NFW profile were also analyzed in \cite{weniger}, based
on 3 years of {\it Fermi}-LAT data. 
Depending on the DM mass, annihilation cross sections down to 
$5\times 10^{-25}$ cm$^3$ s$^{-1}$ can be constrained.
When DM substructures down to the Earth mass, $10^{-6} M_{\odot}$, are present, these
limits could improve possibly going down to the thermal cross section
of $3\times 10^{-26}$ cm$^3$ s$^{-1}$ for DM masses $\leq 150$ GeV and annilation
into $b\bar b$.
In most cases the combined limits are at the level of the strongest individual limit,
unlike the case of dSphs where they can improve the cross-section limits 
a factor of several for a wide range of DM masses.

A similar conclusion was obtained in \cite{ando}, where
49 galaxy clusters were analyzed.
They showed that the stacking analysis only improve 
the cross-section limits by about 10\% at tens of GeV mass regime, and
at most by a factor of $\sim$ 2 at multiple mass regime.
On the other hand, in \cite{ando} a special attention to
modeling DM with baryonic compression was paid.
Although substructure boost was also included in the analysis, the authors attacked this issue in a conservative way, emphasizing that the consideration of Earth-size DM halo requires
extrapolation of the subhalo mass function by 12 orders of magnitude, since present numerical simulations can resolve halos of only about $5\times 10^{7} M_{\odot}$.
They concluded that the substructure boost could exceed the contraction boost, only if the
minimum subhalo mass is considerably smaller than one solar mass.
Otherwise, it can be neglected with respect to the contraction effect.
The authors decided to concentrate the analysis on the Fornax
cluster, which found to be the most promising one yielding the
largest annihilation signal.
This is because of its proximity to the Earth (20 Mpc) and its
relatively large mass ($\sim 10^{14} h^{-1} M_{\odot}$), also
because it does not host any bright active galactic nuclei unlike the
Virgo cluster (that host M87), and finally because it has
regular thermal gas profiles and a spherical central massive elliptical
galaxy, NGC 1399, making the calculation of compression based on the assumption
of spherical symmetry reasonable.
The result was that the annihilation signatures are boosted 
by a factor of 4 due to the compression.
Thus for DM mass of about 10 GeV, the
upper limit, $10^{-25}$ cm$^3$ s$^{-1}$, was obtained.

Unlike above analyses of galaxy clusters, in \cite{frenk} evidence
for extended $\gamma$-ray emission from the Virgo, Coma, and Fornax clusters
based on 3 years of {\it Fermi}-LAT data was 
reported.
When interpreted as annihilation DM particles the data are reproduced
with a particle mass in the range 20--60 GeV annihilating into the
$b\bar b$ channel, or in the range 2--10 GeV and $>$ 1 TeV annihilating into  
$\mu^+\mu^-$. These results seem to be consistent with those obtained in \cite{hooper0} for the galactic center.
The significance found is 4.4, 2.3 and 2.1$\sigma$ for Virgo, Coma and Fornax, respectively, and NFW profiles with substructures were used.
However, recently, Bloom representing the {\it Fermi}-LAT collaboration \cite{bloom} also obtained null results using 
{\it Fermi} data, as previous analyses \cite{weniger,ando}.



\section{Indirect SuperWIMP Searches}
\label{section 4}


The gravitino as the LSP in $R$-parity violating models was first 
studied \cite{yamaguchi} adding the following
bilinear terms in the superpotential of the MSSM:  
$\mu'_i  L_iH_2$. Then,
the detection of gravitino DM through its $\gamma$-ray decay products has been studied 
in the literature \cite{yamaguchi,buchmuller}.
As mentioned in the Introduction, 
unlike other $R$-parity violating models which do not try to address the $\mu$ problem (actually in the bilinear model the problem is augmented with the three new bilinear terms $\mu'_i$), the $\mu\nu$SSM solves 
it and accounts for light neutrino masses.
Thus it seems to be important to know its predictions concerning gravitino DM detection.

In recent works \cite{Choi:2009ng,german}, gravitino DM and its possible detection in the {\it Fermi} satellite when decaying in the galactic halo or extragalactic regions such as the Virgo cluster, were discussed in the
context of the $\mu\nu$SSM.

Summarizing, it was found that a gravitino DM with a mass range of 
0.6--2 GeV, and with a lifetime 
range of about 
$3\times 10^{27}$--$2\times10^{28}$ s would be
detectable by the {\it Fermi}-LAT with a signal-to-noise ratio larger than 3, in 5 years of observations of the Virgo cluster.
On the other hand, gravitino masses larger than about 4 GeV are disfavored 
in the $\mu\nu$SSM by the non-observation of monochromatic lines in the {\it Fermi}-LAT data of the galactic halo. For more details of the computation, where $N$-body simulations of the nearby extragalactic Universe were used, see the talk by G.A. G\'omez-Vargas in these proceedings.

\vspace{0.5cm}

\noindent {\bf Acknowledgments} 

\noindent 
This work was supported by the Spanish MINECO's Consolider-Ingenio 2010 
Programme under grant MultiDark CSD2009-00064, by MINECO under grants 
FPA2009-08958 and FPA2009-09017, by the Comunidad de Madrid under grant 
HEPHACOS S2009/ESP-1473, and by the European Union under the Marie 
Curie-ITN program PITN-GA-2009-237920.


\bibliographystyle{elsarticle-num}

%




\end{document}